\documentclass[
 reprint,
superscriptaddress,
 amsmath,amssymb,
 aps,
superscriptaddress,
 amsmath,amssymb,
 aps,longbibliography
]{revtex4-1}

\usepackage{graphicx}
\usepackage{dcolumn}
\usepackage{bm}
\usepackage{amsmath,amssymb,amstext,amsthm}
\usepackage{braket}
\usepackage{bbold}
\usepackage{bbm}
\usepackage{xcolor}
\usepackage{ gensymb }
\usepackage{soul}
\setstcolor{red}
\usepackage{ mathrsfs }
\usepackage{cancel}
\usepackage{esvect}

\usepackage{hyperref}

\hypersetup{colorlinks,linkcolor=blue,citecolor=blue,urlcolor=blue}

\newcommand{\pend}{pendell\"osung }

\usepackage{etoolbox}

\AtBeginEnvironment{equation}{\vspace{6pt}}

\begin{document}

\preprint{APS/123-QED}
\setlength{\abovedisplayskip}{1pt}
\title{A perfect crystal neutron loop cavity}

\author{Owen Lailey} 
\email{oalailey@uwaterloo.ca}
\affiliation{Institute for Quantum Computing, University of Waterloo,  Waterloo, ON, Canada, N2L3G1}
\affiliation{Department of Physics and Astronomy, University of Waterloo, Waterloo, ON, Canada, N2L3G1}

\author{Dusan Sarenac}
\affiliation{Department of Physics, University at Buffalo, State University of New York, Buffalo, New York 14260, USA}

\author{David G. Cory}
\affiliation{Institute for Quantum Computing, University of Waterloo,  Waterloo, ON, Canada, N2L3G1}
\affiliation{Department of Chemistry, University of Waterloo, Waterloo, ON, Canada, N2L3G1}

\author{Michael G. Huber}
\affiliation{National Institute of Standards and Technology, Gaithersburg, Maryland 20899, USA}

\author{Dmitry A. Pushin}
\email{dmitry.pushin@uwaterloo.ca}
\affiliation{Institute for Quantum Computing, University of Waterloo,  Waterloo, ON, Canada, N2L3G1}
\affiliation{Department of Physics and Astronomy, University of Waterloo, Waterloo, ON, Canada, N2L3G1}

\date{\today}

\pacs{Valid PACS appear here}

\begin{abstract}
    Coherent control of neutrons via Bragg diffraction forms the foundation of perfect crystal neutron interferometry, facilitating both fundamental tests of quantum mechanics and applications in quantum information science. In cavity geometries, perfect crystals enable neutron confinement and have been employed in precision measurements of spin-orbit interactions and for neutron electric dipole moment (nEDM) searches. However, in these conventional configurations, neutrons undergo a single pass through the crystal geometry, placing a physical constraint on both crystal and in-flight interaction times and measurement sensitivity. In this work, we introduce a neutron loop cavity that coherently recirculates neutrons through repeated Bragg reflections between perfect silicon crystal blades. This structure is predicted to achieve a neutron survival probability of $\sim64~\%$ for 10,000 Bragg reflections, corresponding to confinement times on the order of seconds. We propose a Schwinger interaction measurement that achieves a $\pi$ spin rotation in 800 Bragg reflections, representing more than a tenfold improvement in sensitivity over recent measurements. Further applications include high-sensitivity nEDM searches targeting the $10^{-27}$~e$\cdot$cm scale, as well as competitive experimental tests of neutron parity violation, the neutron lifetime, and the quantum Zeno effect with neutrons.
    
\end{abstract}
\maketitle

\section{Introduction}
Perfect crystal neutron interferometers (NIs) have been instrumental in experimentally verifying foundational quantum phenomena~\cite{Sam}. Landmark experiments include the confirmation of the $4\pi$ spin rotation symmetry of spin-$1/2$ particles~\cite{Rauch_1975_PhysLetta} and the observation of gravitationally induced phase shifts in neutron matter waves~\cite{colella1975observation}. More recently, NIs have enabled studies of decoherence-free subspaces for quantum error correction~\cite{decoherence}, hypothetical fifth forces~\cite{li2016neutron, heacock2021pendellosung}, quantum Cheshire cat and operator commutation relations~\cite{cheshire_cat}, and neutron orbital angular momentum (OAM) states~\cite{clark2015controlling, sarenac2016holography, geerits2023phase}.

\begin{figure}[h]
    \centering
    \includegraphics[width=1\linewidth]{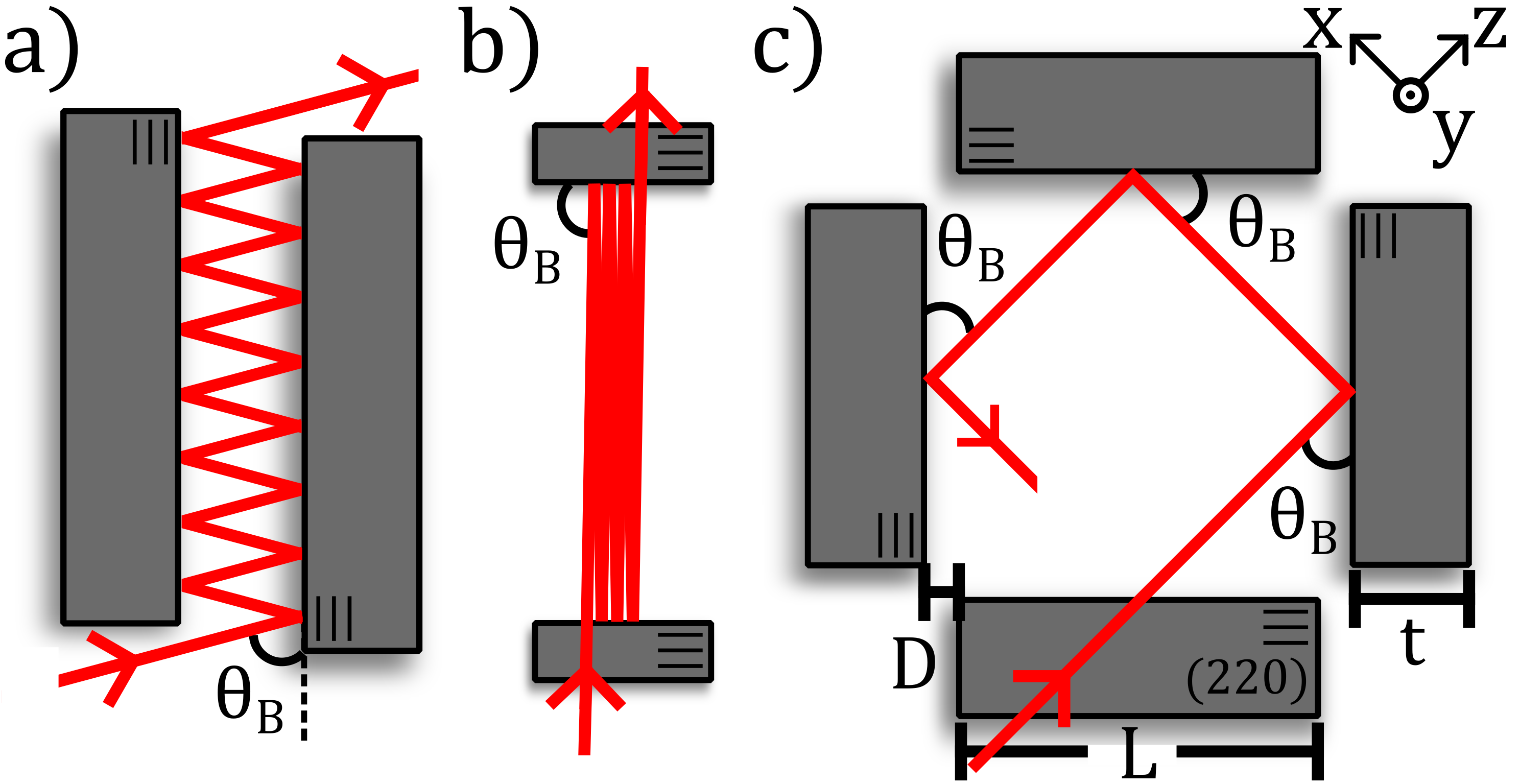}
    \caption{Schematic of neutron propagation through three perfect crystal cavity geometries: (a) compact double Bragg blade~\cite{gentile, neutron_cav}, (b) extended double Bragg blade~\cite{vesta}, and (c) the proposed loop cavity. In (a-b), the crystals are attached to a common base and $\theta_B$ is near $90\degree$ to maximize the number of bounces. In (c), the four identical Bragg blades are independently mounted and aligned to operate at $\theta_B = 45\degree$, confining neutrons to repeatedly traverse the same path. Note that only a few pathways of diffracted intensity are shown for simplicity.}
    \label{fig:NIs}
\end{figure}

In parallel with neutron interferometry, perfect-crystal neutron cavities formed by two Bragg-diffracting silicon blades have been developed to confine and coherently control neutron beams~\cite{schuster1990test, schuster1992cold, jericha1996performance, BB_storage, jaekel1999new, vesta, gentile, dombeck2001measurement}. These devices (see Fig.~\ref{fig:NIs}a,b) have demonstrated hundreds to thousands of Bragg reflections at high Bragg angles near $90\degree$, extending effective interaction times and enhancing sensitivity to subtle quantum effects. This capability has enabled direct observation of neutron spin rotation from the spin-orbit interaction in silicon using $\sim100$ Bragg reflections~\cite{gentile}, as well as high-precision reflectivity measurements relevant to neutron electric dipole moment (nEDM) searches, where the measured reflectivity corresponds to losses low enough to permit up to $\sim20{,}000$ reflections~\cite{dombeck2001measurement}. However, in these double Bragg neutron cavities, both crystal and in-flight interaction times are ultimately limited either by the finite device geometry, which restricts the number of achievable reflections in compact setups~\cite{gentile}, or by practical constraints such as neutron guide lengths, beam divergence, and alignment stability when the cavity is extended to meter scales~\cite{vesta}. This trade-off motivates the development of closed-loop neutron recirculation, which enables repeated passes within a compact device.

Leveraging recent advances in NI fabrication~\cite{huber2024achieving}, demonstrated high-precision alignment of Bragg blades in split-crystal interferometers~\cite{lemmel2022neutron}, and versatile dynamical diffraction modeling~\cite{Nsofini_2016, nsofini2017noise, nsofini2019coherence, nahman2022generalizing, neutron_cav}, we present a novel neutron loop cavity design that confines neutrons through a repeated four Bragg diffractions in a closed trajectory. This geometry enables controlled amplification of small, cumulative interactions. We estimate that a confinement probability of approximately $64~\%$ can be achieved after $10^4$ Bragg reflections, where the confinement probability is defined as the fraction of neutrons remaining in the cavity after $N$ reflections. As a near-term application, we propose a simplified measurement of the Schwinger spin-orbit interaction, in which the looped neutron trajectory produces a $\pi$ spin rotation within approximately 800 reflections. This unique setup may help resolve the $\sim 40~\%$ discrepancy between the recent experimental Schwinger measurements and theoretical predictions~\cite{gentile}. Furthermore, we describe how this device may be utilized for competitive nEDM searches, neutron parity violation measurements, neutron lifetime measurements, and tests of the quantum Zeno effect with neutrons.
\vfill

\section{Dynamical diffraction modeling}
For the design and analysis of the loop cavity we apply the quantum information (QI) model for dynamical diffraction (DD) theory~\cite{Nsofini_2016, nsofini2017noise, nsofini2019coherence, nahman2022generalizing, neutron_cav}. The QI model describes neutron propagation through perfect crystals as a quantum random walk through a lattice of nodes. Each node is a unitary operation on the neutron state which updates the neutron wavefunction at each point inside the crystals. The QI model has been shown to successfully describe the beam intensity profiles in a conventional neutron interferometer~\cite{Nsofini_2016}, as well as neutron confinement in a perfect crystal cavity composed of two Bragg blades~\cite{neutron_cav} (see Fig.~\ref{fig:NIs}a). Complete description of the model as a quantum random walk and the equivalence with DD equations can be found in Refs.~\cite{Nsofini_2016, nsofini2017noise, nsofini2019coherence, nahman2022generalizing, neutron_cav}. 

\indent In DD, the characteristic length scale is the \pend length which is on the order of $50~\mu$m~\cite{Sam}. As neutrons propagate through perfect crystal blades, intensity oscillates back and forth between the transmitted and reflected beams with a period equal to the \pend length:
\begin{equation}
    \Delta_H = \frac{\pi V_{cell}}{\lambda |F_H|}
    \begin{cases}
    \cos{\theta_B} & \text{Laue} \\
    \sin{\theta_B} & \text{Bragg}
    \end{cases}
    \label{eq:Pend}
\end{equation}
where $V_{cell}$ is the volume of the crystal unit cell, $\lambda$ is the incident neutron wavelength, $F_H$ is the crystal structure factor, and $\theta_B$ is the Bragg angle. For $\lambda = 2.72$ \AA~incident on a perfect silicon crystal (220), $\theta_B \approx 45\degree$ and $\Delta_H \approx 40~\mu$m. Unless otherwise noted, these values are used for all quantities and calculations throughout the text.

Neutron storage devices that utilize Bragg diffraction exploit the fact that neutrons within a narrow angular range (or equivalently momentum) about the Bragg condition reflect with unity probability. This range of total reflection, known as the Darwin width, is given by:
\begin{equation}
    \delta\theta = \frac{|F_H|}{\pi V_{cell}}\frac{\lambda^2}{\sin(2\theta_B)},
    \label{eq:darwin}
\end{equation}
which gives $\delta\theta \approx 5~\mu$rad. The Darwin width can be nearly doubled for $\lambda = 4.4$ \AA~and (111), while keeping $\theta_B \approx 45\degree$.

\section{Neutron loop cavity design}
\label{design}

\begin{figure*}[ht]
    \centering
    \includegraphics[width=1\linewidth]{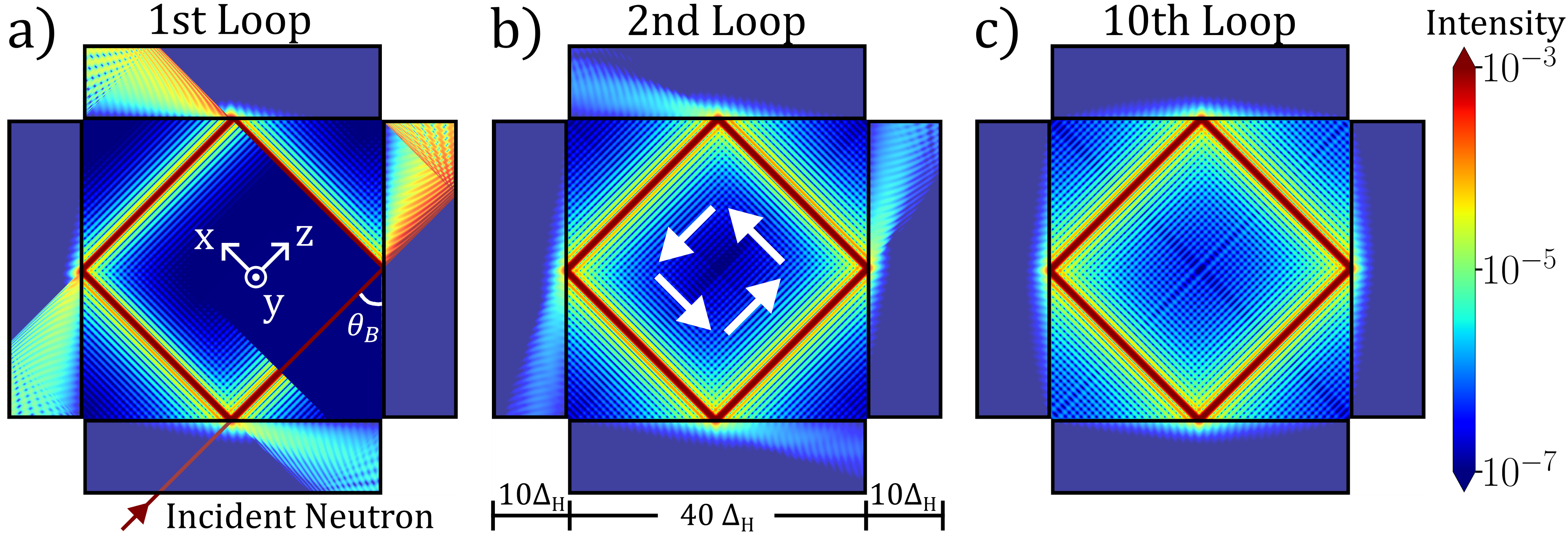}
    \caption{(a) Simulated neutron intensity inside the loop cavity after the first four reflections. Neutrons are incident to the rightmost Bragg blade at $\theta_B = 45\degree$, undergoing subsequent reflections in a counter clockwise direction. As the number of reflections within the loop increases (b-c), the neutrons within the Darwin width remain confined, with minimal intensity escaping through the crystals. In these simulations, the crystal thickness and crystal length are compact for visualization purposes.}
    \label{fig:2d}
\end{figure*}

A schematic of the proposed loop cavity is illustrated in Fig.~\ref{fig:NIs}c. Four identical Bragg blades of length $L$ and thickness $t$ are positioned independently in a square formation. The crystal separation distance can be increased beyond the length $L$ by introducing a gap distance $D$ between all crystals. Neutrons are incident upon the rightmost blade at $\theta_B = 45^\circ$. Subsequent reflections from the blades confine the neutron to a closed, circulating trajectory. This cavity geometry requires $\theta_B = 45^\circ$, ensuring that the reflection from one blade is incident upon the next at $90^\circ-\theta_B = \theta_B$. Neutrons within the Darwin width about the Bragg angle are accepted and undergo repeated reflections. Similar designs have been considered for X-ray resonators~\cite{bond1967proposed, deslattes1968x, cotterill1968universal, liu2024x}. Alignment conditions, along with experimental requirements such as crystal thickness tolerances and loading/unloading of the cavity, are discussed in section~\ref{requirements}.

\subsection{Confined neutron wavefunction}
The central purpose of a neutron cavity is to confine neutrons for an extended amount of time with a high survival probability. We analyze the propagation of the neutron wavefunction within the loop cavity as a function of the number of reflections $N$ using the QI model. Figure~\ref{fig:2d} shows the evolution of the neutron probability intensity within the crystal blades and surrounding space for $t = 10\Delta_H$ and $L = 40\Delta_H$. Here the simulated crystal dimensions are kept compact to clearly visualize the intensity inside the blades and within the loop. Neutrons transmit through the bottom Bragg blade without diffracting such that the first Bragg reflection will occur at the rightmost Bragg blade (discussed in loading/unloading~\ref{requirements}). At each subsequent coherent interaction, the transmitted component escapes while the reflected component remains confined; four reflections constitute one complete closed-loop.

Figures~\ref{fig:2d}b and \ref{fig:2d}c show the neutron intensity after the second and tenth completed loops. By this stage, transmission through the Bragg crystals is negligible. This behavior is consistent with DD theory: the cavity acts as a filter, retaining only those neutrons with wavelengths within the Darwin width. For these confined neutrons, the reflectivity of the Bragg crystals is nearly unity.

\begin{figure*}[ht]
    \centering
    \includegraphics[width=1\linewidth]{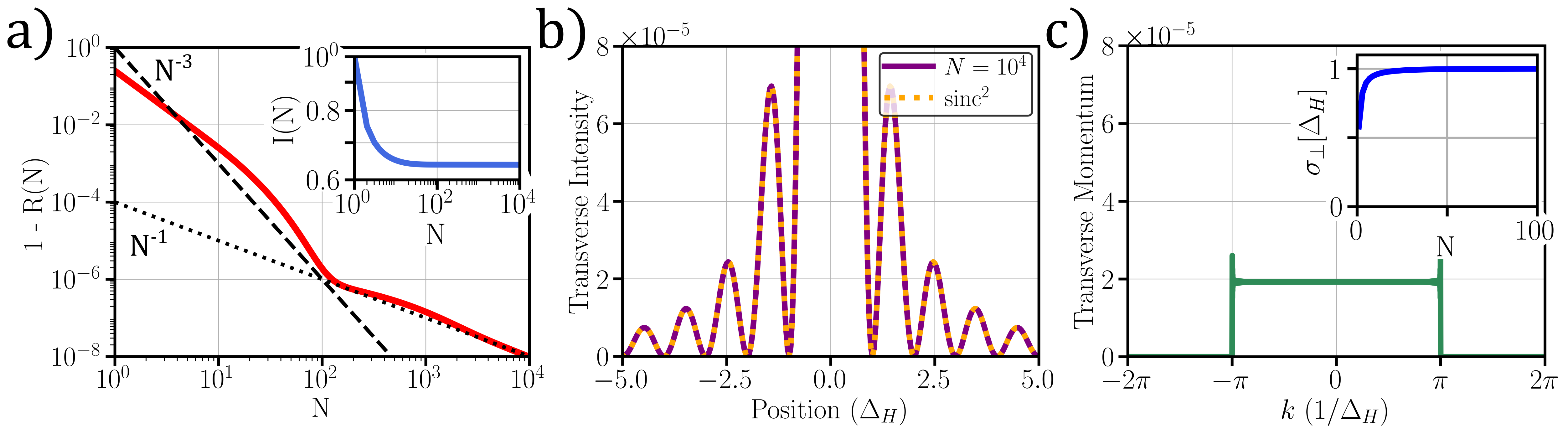}
    \caption{Simulation of the designed cavity performance ($t = 100\Delta_H$, $L = 2000\Delta_H$). (a) The evolution of the reflectivity, plotted as ($1-R(N)$), versus the number of reflections $N$. The reflectivity reaches $R = 1-10^{-8}$ at $N = 10^4$ reflections. The escape probability ($1-R(N)$) follows two distinct power-law regimes: an initial rapid decay ($\propto N^{-3}$) transitioning to a shallower dependence ($\propto N^{-1}$), indicated by the dashed lines. The inset demonstrates the confined intensity $I(N)$ versus the number of reflections $N$. (b) Transverse intensity profile after $10^4$ reflections, which is fit to a $\mathrm{sinc}^2$ function. (c) Corresponding transverse momentum distribution showing the rectangular pulse characteristic of the Darwin plateau, with visible Gibbs ringing. The inset demonstrates the neutron transverse coherence $\sigma_\perp$ rapidly saturating to $1~\Delta_H$ as a function of $N$, in agreement with the confined neutron momentum becoming restricted to the Darwin plateau. }
    \label{fig:main}
\end{figure*}

This behaviour is quantified in Fig.~\ref{fig:main}a, which plots the confined intensity for up to $N = 10^4$ reflections ($t=100\Delta_H,~L=2000\Delta_H$). The inset shows the deviation of the single-bounce reflectivity from unity, defined as $R(N) = I(N) / I(N-1)$, where $I(N)$ is the integrated intensity of the $N$-th reflected wave over the entire inner crystal face. The reflectivity follows two distinct regimes, reaching $R(10^4) = 1 - 10^{-8}$. Initially, the escape probability $(1-R(N))$ decreases rapidly ($\propto N^{-3}$) until a transition point $N_{trans}=100$, which scales with the Bragg blade thickness as $N_{trans} \sim t^2$. This early behavior arises from the finite crystal thickness, which supports complex modes involving multiple internal reflections. Eventually, the dominant neutron mode stabilizes to the main reflection from the inner Bragg face. The escape probability then transitions to a shallower dependence ($\propto N^{-1}$). The cumulative survival probability is given by $P = \prod_{i=1}^N R(i)$. Consequently, neutrons satisfying the Bragg condition have a $P \approx 64~\%$ chance of surviving $10^4$ reflections. 

From DD theory for successive Bragg reflections, the average reflectivity is:
\begin{equation}
    \left<R\right> = 1 - \Delta_e n(\sigma_{abs} + \sigma_{therm}),
\end{equation}
where $\Delta_e$ is the wavefunction extinction depth, $n$ is the atomic density, $\sigma_{abs}$ is the absorption cross section, and $\sigma_{therm}$ is the diffuse scattering cross section~\cite{dombeck2001measurement}. The extinction depth, where the amplitude is reduced by $e^{-1},$ is defined as $\Delta_e = \sin{\theta_B}e^{-W} / (\lambda n b_c)$, where $e^{-W}$ is the Debye-Waller factor, and $b_c$ is the neutron coherent scattering length for silicon. Using $e^{-W} = 0.96$, $\sigma_{abs} = 0.26$~barn, and $\sigma_{therm} = 0.161$~barn, the theoretical average reflectivity is $\left<R\right> = 0.999975$. The survival fraction is then given by: $\eta = \epsilon \text{exp}^{-N(1-\left<R\right>)}$, where $\epsilon = 0.9$ is an efficiency factor taking into account the sharp decrease in reflectivity near the edge of the Darwin plateau~\cite{dombeck2001measurement}. For $N = 10^4$ reflections, DD theory predicts a survival probability of $\eta \approx 70~\%$, in approximate agreement with simulations.

Figure~\ref{fig:main}b shows the confined transverse intensity profile as a function of position across the inner reflecting crystal face after $10^4$ reflections, resembling a $\mathrm{sinc}^2$ distribution. The corresponding momentum distribution forms a rectangular pulse (Fig.~\ref{fig:main}c). This agrees with DD theory, confirming that only neutrons with momenta strictly within the Darwin width survive the extended confinement. Directly related to this phenomena, we consider the neutron transverse coherence $\sigma_\perp$ as a function of the number of reflections $N$, where we set $\sigma_\perp$ as the first zero of the sinc$^2$ intensity distribution. The transverse extent rapidly saturates to $1~\Delta_H$ according to $\sigma_\perp \sim 1 - \frac{1}{2}e^{-N/3}$.

\subsection{Experimental design requirements}
\label{requirements}
We outline key experimental considerations for implementing the loop cavity.\\

\noindent \textit{Crystal alignment:}\\
\indent To ensure near-unity reflectivity, each Bragg blade requires alignment to each other within the Darwin width (Eq.~\ref{eq:darwin}). Consistent with established methods in neutron and X-ray optics~\cite{bonse1968two, lemmel2022neutron}, the crystals are mounted on piezo stages for fine rotational and translational adjustment. For instance, the recently demonstrated split-crystal neutron interferometer utilized a piezo-driven tip–tilt platform achieving a $\pm70~\mu$rad range with sub-nanoradian resolution~\cite{lemmel2022neutron}. The proposed loop cavity employs independently mounted piezo rotation stages with sub-$\mu$rad resolution and high resonant frequencies, enabling $\mu$rad-scale corrections on sub-millisecond time scales.\\

\noindent \textit{Loading/unloading:}\\
\indent Several crystal gating techniques have been experimentally demonstrated in BB-type cavities for cold neutron storage~\cite{schuster1990test, vesta}. The task of loading/unloading is to temporarily shift incident neutrons off the Bragg condition by the Darwin width $\delta\theta$ (Eq.~\ref{eq:darwin}). Since the Bragg crystals are independently mounted on piezo-driven rotation stages for alignment, one crystal can be rotated by an angle $\delta\theta$ to load the device and subsequently returned by $-\delta\theta$, restoring the Bragg condition and confining the neutrons within the loop. For $\lambda = 2.72~\text{\AA}$ and a crystal length $L = 2000\Delta_H$, a neutron completes a single round trip in $\sim 0.15~\text{ms}$. The cavity can be continuously loaded with pulses of width $\delta t = \frac{L/2 + t}{\sqrt{2}v}$, where $v$ is the neutron velocity. Including a separation $D$ between the crystals will increase the neutron round-trip time, which correspondingly relaxes the bandwidth requirements on the piezo-driven rotation stages used to restore the Bragg condition.

Another convenient option leverages the neutron’s magnetic moment, which couples to an external magnetic field, producing a Zeeman energy shift and a corresponding change in the neutron wavevector:
\begin{equation}
    \Delta k=\pm\frac{\mu B m}{\hbar^2k},
\end{equation}
where $\mu$ is the neutron magnetic moment, $B$ is the applied magnetic field, $m$ is the neutron mass, and $k$ is the neutron wavevector. A magnetic field pulse of approximately 3.6~T produces a wavevector shift equivalent to the Darwin width, enabling the neutron to transmit through the crystal. This method provides a means of both loading and unloading neutrons into and out of the loop cavity~\cite{schuster1990test}.\\

\noindent \textit{Gravity:}\\
\indent To maintain neutron confinement within the loop cavity, gravitational effects must be compensated. As demonstrated in other neutron storage systems~\cite{BB_storage}, this can be achieved by incorporating neutron guides between the crystal blades, allowing falling neutrons to reflect from the lower guide surface with near-unity probability. In the loop cavity geometry, four neutron guides are required, one for each path segment between adjacent blades. Additionally, these guides confine the beam in the vertical direction, thereby limiting its divergence. Since the meter-long guides implemented in earlier neutron storage systems~\cite{BB_storage, vesta}, neutron guide technology has advanced substantially through the adoption of multilayer supermirror coatings, higher-m reflectors, and optimized guide geometries, significantly improving neutron transport efficiency and delivered flux at modern facilities~\cite{le2023upgrade, nist_guide_upgrade, abele2006characterization}. For short confinement times, it may be feasible to dynamically lower the cavity, eliminating the need for guides while still accounting for the neutron free fall.\\

\begin{figure}[t]
    \centering
    \includegraphics[width=0.9\linewidth]{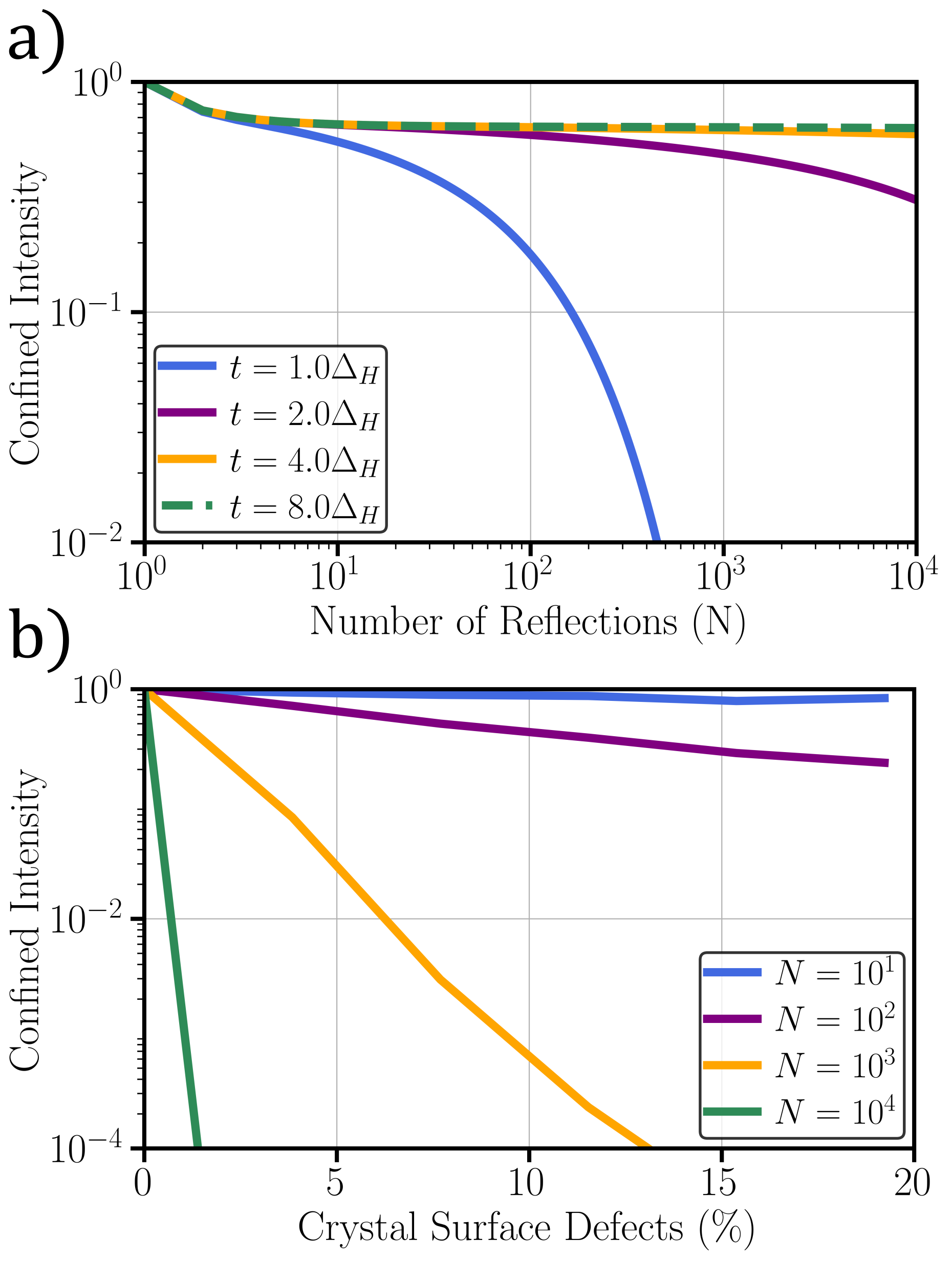}
    \caption{(a) Confined intensity as a function of the number of reflections $N$ for various Bragg blade thicknesses $t$ and $L = 2000\Delta_H$. For $t < 4\Delta_H$, intensity loss increases with $N$, while for $t \geq 4\Delta_H$, confinement is maintained and does not significantly improve with further increases in thickness. (b) Confined intensity as a function of crystal blade surface roughness/defect density. Sensitivity to surface imperfections increases with the number of reflections. For up to $10^2$ reflections, roughness has minimal impact, but for $10^4$ reflections, crystal surfaces must be defect-free to preserve confinement.}
    \label{fig:params}
\end{figure}

\noindent \textit{Crystal thickness and surface defects:}\\
\indent The neutron confinement probability as a function of Bragg blade thickness $t$ is shown in Fig.~\ref{fig:params}a. To sustain confinement over thousands of reflections, the Bragg blades must have a thickness of at least $t \geq 4\Delta_H$.

Surface roughness and crystal defects are expected to limit neutron confinement~\cite{neutron_cav, authier2012x}. Within the QI model, surface roughness can be incorporated by randomly replacing small regions of crystal within one $\Delta_H$ of the surface with empty space. This approach follows the methodology of Ref.~\cite{neutron_cav}, which modeled imperfections in an experimental BB cavity. Figure~\ref{fig:params}b presents the confined neutron intensity versus reflection number $N$ for both an ideal defect-free cavity and cavities with surface roughness on the Bragg blade faces. For up to $10^2$ reflections, surface roughness has little effect on confinement. At larger $N$, however, its impact becomes pronounced: at $10^3$ reflections, a surface defect density of approximately $10~\%$ leads to substantial intensity loss, while for $10^4$ reflections, perfect crystal surfaces are required to maintain confinement. This heightened sensitivity arises from the repeated interaction of the confined neutron wavefunction with the same crystal surface regions, in contrast to conventional BB cavities, where neutrons interact with each crystal surface only once.

\section{Fundamental Physics Applications}
\label{apps}

Here we outline several promising experimental applications, including measurements of the neutron spin-orbit interaction, neutron electric dipole moment, nucleon-nucleon weak interaction, neutron lifetime, and neutron quantum Zeno effect, based on first-order estimates of confinement time, phase accumulation, and interaction strength. Emphasis is placed on contrasting the unique loop cavity geometry with previous experimental implementations and proposals. Detailed systematics and statistical comparison to current experimental approaches will be addressed in future work following experimental construction and characterization of the neutron loop cavity.

Considering practical cold, monochromatic neutron beams with flux on the order of $10^5\ \mathrm{n}/(\mathrm{cm}^2\cdot\mathrm{s})$, a beam size of $1~\mathrm{cm}^2$ corresponds to an incident rate of $10^5$ neutrons/s at the cavity entrance. For a beam divergence of $1~\mathrm{mrad}$ and a Darwin width of $5~\mu\mathrm{rad}$, only an approximate fraction $\sim5 \times 10^{-3}$ of the incident neutrons are accepted, yielding a rate of $500$ neutrons/s. As discussed in section~\ref{design}, the survival probability after $10^4$ reflections is approximately $64\%$, corresponding to $\sim 300$ neutrons/s exiting the cavity and reaching the detector. Accounting for additional reductions due to crystal misalignment, neutron transport losses, timing inefficiencies in loading and unloading, and the inclusion of spin polarizers and analyzers, the expected detected count rate is conservatively on the order of $1-10~\mathrm{Hz}$.

\subsection{Neutron spin-orbit interaction and nEDM}
The neutron spin-orbit interaction, or Schwinger scattering, induces a small spin rotation of the neutron due to the coupling of its magnetic dipole moment (MDM) with the atomic electric field during a single Bragg reflection~\cite{schwinger1948polarization, shull1963neutron}. This interaction has been utilized in past efforts to search for the nEDM using Bragg reflection from perfect crystals~\cite{shull1967search}. More recently, the Schwinger interaction in noncentrosymmetric perfect crystals has been experimentally explored as a basis for nEDM measurements~\cite{fedorov2006neutron, fedorov2010measurement, fedorov2018modern}. To enhance sensitivity to either the spin-orbit interaction or the nEDM in Bragg scattering, several techniques have been proposed that involve hundreds to thousands of Bragg reflections within slotted crystals to increase the neutron's interaction time with the internal electric field~\cite{gentile, dombeck2001measurement, zeilinger1984symmetry}. The loop cavity enables orders of magnitude improvement in the interaction time, along with other practical advantages, potentially enabling higher sensitivity Schwinger and nEDM measurements.

The motion of a neutron through the atomic electric field $\vec{E}$ induces an effective magnetic field in the neutron rest frame:
\begin{equation}
    \vec{B}_{eff} = \frac{1}{c}\vec{v} \times \vec{E},
\end{equation}
where $c$ is the speed of light and $\vec{v}$ is the neutron velocity. In this field, a neutron with spin polarization $\vec{P}$ will rotate around $\vec{B}_{eff}$ due to the torque on the MDM:
\begin{equation}
    \vec{\tau}_{MDM} = \mu_n\vec{P}\times \vec{B}_{eff},
\end{equation}
where $\mu_n$ is the nuclear magnetic moment. The expected magnitude of rotation for a single Bragg reflection is approximately:
\begin{equation}
    \theta_i = 2|b_{so}/b_c|,
\end{equation}
where $b_{so}$ is the Schwinger scattering length and $b_c = 0.415\times10^{-12}$~cm is the coherent nuclear scattering length for silicon. For $\theta_B = 85\degree$ and (220) Bragg reflection used in Ref.~\cite{gentile}, the rotation angle is $\theta_i = 3.27\times 10^{-4}$~rad. By employing a slotted crystal, 136 reflections were achieved, magnifying the spin rotation to a measured $4.73\times10^{-2}$~rad. Notably, for the Schwinger rotation to add constructively in the BB geometry, the entire crystal was placed in a uniform magnetic field such that the neutron spin undergoes a precession of angle $\pi$ between successive Bragg reflections from opposing crystal faces. In the absence of this condition, the Schwinger spin rotation does not accumulate with each reflection as the effective magnetic field $\vec{B}_{eff}$ reverses sign at each reflection. In Ref.~\cite{gentile}, the measured spin rotation was $40~\%$ larger than expected, indicating a need for further characterization of the Schwinger interaction with an improved setup.

The loop cavity geometry supports a large number of Bragg reflections without requiring the high Bragg angles characteristic of slotted crystal designs. For the (220) reflection at $\theta_B = 45^\circ$, the Schwinger rotation per reflection is $\theta_i = 3.96 \times 10^{-3}\,\mathrm{rad}$, representing an order-of-magnitude enhancement compared to standard configurations. Moreover, the condition for constructive accumulation of the Schwinger rotation is simplified due to the loop geometry. The $90^\circ$ rotational symmetry of the cavity ensures that $\vec{v}\times\vec{E}$ points in the same direction for each Bragg reflection. Consequently, the effective magnetic field $\vec{B}_{\mathrm{eff}}$ has the same orientation at every Bragg blade. This contrasts with double-Bragg cavities, where $\vec{B}_{\mathrm{eff}}$ reverses sign between reflections~\cite{gentile}. A guide field, however, is still required to preserve the neutron polarization: it maintains the polarization along $(x+z)/\sqrt{2}$ throughout the device (see Fig.~\ref{fig:2d}), and should be tuned to avoid partial cancellation as the spin precesses between reflections. The accumulated spin rotation reaches $0.54$~rad after 136 reflections and a full $\pi$~rad after 800 reflections, nearly 100 times larger than previous measurements~\cite{gentile}. 

In Ref.~\cite{dombeck2001measurement}, a slotted crystal with reflectivity $R \geq 0.99998$ was reported, possibly enabling tens of thousands of bounces. The authors estimated that by coupling the neutron EDM to the atomic electric field at each Bragg reflection~\cite{shull1967search} a sensitivity of order $10^{-27}\,e\cdot\mathrm{cm}$ could be achievable, potentially surpassing the current experimental constraint $d_n = (0.0\pm 1.1_{\text{stat}} \pm 0.2_{\text{sys}})\times10^{-26}\,e\cdot\mathrm{cm}$~\cite{abel2020measurement}. For any nEDM measurement, the statistical uncertainty is proportional to the product $E\tau N^{1/2}$, where $E$ is the electric field, $\tau$ is the interaction time, and $N$ is the number of counted neutrons~\cite{lamoreaux2009experimental}. Compared to the slotted crystal estimate, the loop cavity is predicted to achieve comparable or better reflectivity and hence increased interaction times. Thus, a measurement of the nEDM with sensitivity of $10^{-27}\,e\cdot\mathrm{cm}$ may be possible with the loop cavity, repositioning neutron nEDM searches based on crystal diffraction as a competitive method compared to current best measurements with ultracold neutrons~\cite{abel2020measurement}.

\subsection{Parity violation with liquid Helium-4}
The nucleon-nucleon (NN) weak interaction remains a poorly constrained sector of the Standard Model~\cite{schindler2013theory}. Parity violation (PV) serves as a unique filter to isolate the weak force from the dominant, parity-conserving strong interaction, which typically suppresses PV amplitudes by a factor of $10^{-6}$--$10^{-7}$. To theoretically interpret these effects, few-nucleon systems, such as with neutrons, are essential; they allow a straightforward mapping between experimental observables and underlying NN interactions~\cite{schindler2013theory}. In liquid helium-4, this interaction manifests as a rotation of the neutron spin about its momentum vector, theoretically predicted to be on the order of $10^{-6}$--$10^{-7}$~rad/m. The current best experimental limit, $d\phi/dz = [+1.7 \pm 9.1(\mathrm{stat}) \pm 1.4(\mathrm{sys})] \times 10^{-7}\,\mathrm{rad/m}$~\cite{snow2011upper}, is dominated by large statistical uncertainties.

The loop cavity may provide a $2\times$ improvement in statistical sensitivity by enabling optimal $^4$He interaction lengths, while simultaneously using a simplified sample environment and distinct systematic controls. In Ref.~\cite{snow2011upper}, two $0.42$~m liquid $^4$He targets were alternately filled, with a $\pi$ spin flipper placed between them, in order to isolate the PV neutron spin rotation signal. The use of long targets is motivated by the linear scaling of the PV rotation angle, $\phi \propto \rho z$, with number density of scatterers $\rho$ and target length $z$. However, for transmission experiments, the statistical uncertainty $\propto \sqrt{e^{-\rho \sigma_{tot} z}}$ is minimized when the target length equals twice the neutron mean free path, $z = 2/(\rho \sigma_{\mathrm{tot}})$, where $\sigma_{\mathrm{tot}}$ is the total scattering cross section. For liquid $^4$He at the cold-neutron energies used in the experiment, this optimal length is approximately $2$~m; however, only $0.42$~m targets were used due to the 100~cm maximum bore length of the cryostat~\cite{bass2009liquid}. Achieving an effective interaction length near 2~m would therefore yield a $2\times$ gain in sensitivity. In the loop cavity, this condition can be met by repeatedly circulating neutrons through a much shorter target; for example, a 1~cm liquid $^4$He sample traversed 200 times yields an effective path length of 2~m.

In addition to improving sensitivity, the loop cavity may enable more direct control of parity-conserving backgrounds. In Ref.~\cite{snow2011upper}, background subtraction required alternately filling and draining two liquid $^4$He target chambers with a $\pi$ spin flipper placed between them in order to isolate the PV neutron spin rotation signal. In the loop cavity, however, the sign of the PV spin rotation may be reversed by rotating the entire cavity by $90^\circ$ about the entrance crystal, which reverses the neutron propagation direction through the loop from counterclockwise to clockwise. This transformation flips the neutron momentum at the target location, $\mathbf{k} \rightarrow -\mathbf{k}$. Since the PV rotation changes sign under $\mathbf{k} \rightarrow -\mathbf{k}$, while many parity-conserving magnetic backgrounds do not, the PV signal may be isolated without modifying the target configuration or requiring a spin flipper.

\subsection{Neutron lifetime}
\label{lifetime}
The ``neutron lifetime puzzle'' remains an active area of research~\cite{musedinovic2025measurement, caylor2025detection, fuwa2024improved}, with the mean neutron lifetime $\tau_n$ a key input parameter for models of the early universe~\cite{chowdhury2023neutron}. Two main experimental approaches are used, with ongoing disagreement at the $\sim1~\%$ level. The beam method primarily measures the rate of proton production from neutron beta decay in a cold neutron beam,
\begin{equation}
    n \rightarrow p + e^- + \overline{\nu_e},
\end{equation}
while the bottle method measures the survival of stored ultra-cold neutrons as a function of time. A neutron lifetime measurement using a loop cavity would fall under the bottle method, but with neutron energies $\sim10^5\times$ larger (comparable to beam experiments) and a fundamentally different experimental geometry. Furthermore, the cold neutron flux is many orders of magnitude higher than that available for ultracold neutrons, making the loop cavity a unique variant of a bottle-type experiment. 

Neutrons are loaded into the loop, stored for a variable time $\Delta t$, and then extracted and counted. The measured quantity is the number of surviving neutrons $N(\Delta t)$. In addition to beta decay, neutrons are lost through imperfect reflections from the crystal mirrors. This is a well-known issue in bottle experiments and is typically addressed by varying the trap geometry to change the collision rate. Extrapolating to zero collision rate allows the decay lifetime to be extracted. In the loop cavity, this can be achieved by changing the separation $D$ between independently mounted crystals, which increases the path length and reduces the reflection rate.

The feasibility of the loop cavity depends on if neutrons can be confined for times comparable to or exceeding $\tau_n \approx 15$~min. For a loop length $L = 2000\Delta_H$ (and $D = 0$), a neutron completes one circuit in approximately $0.15$~ms. Over 15 minutes, this corresponds to roughly $6 \times 10^6$ Bragg reflections, requiring reflectivity $R > 1-10^{-6}$ for confinement. Theoretically, such reflectivities are achievable, as demonstrated in Fig.~\ref{fig:main}a, where values as high as $R = 1-10^{-8}$ are shown. Experimentally, reflectivities of $R = 1-2\times10^{-5}$ were reported for slotted silicon crystals~\cite{dombeck2001measurement}, which would yield a confinement time of seconds in the loop cavity. However, given the substantial advances in crystal fabrication and polishing in recent years~\cite{huber2024achieving}, the necessary reflectivity may be within technical reach. Consequently, a neutron lifetime measurement using a loop cavity offers a unique bottle-type measurement with beam-energy neutrons and entirely different systematic uncertainties. We emphasize that the forthcoming experimental implementation of the neutron loop cavity will allow for a detailed comparison against the systematics and counting statistics of existing bottle measurements.

\subsection{Quantum Zeno effect and Zeno dragging}
The quantum Zeno effect (QZE) is a fundamental prediction of quantum mechanics: frequent measurements can suppress the coherent time evolution of a quantum system~\cite{misra1977zeno, itano1990quantum}. It has been demonstrated in many physical systems~\cite{itano1990quantum, fischer2001observation, kwiat1995interaction}, but has yet to be observed with neutrons due to the experimental complexity of existing proposals~\cite{pascazio1993quantum, rauch2001quantum, vesta, facchi2003optimization}. The loop cavity provides a simple and controlled platform to test the QZE with neutrons and also enables the related concept of quantum Zeno dragging, which has applications in quantum control~\cite{lewalle2024optimal, hacohen2018incoherent}.

To illustrate the QZE for neutrons, consider a neutron initially in the spin-up state ($\uparrow$) polarized along the $z$ axis, passing through a magnetic field oriented along the transverse $x$ axis (see Fig.~\ref{fig:2d}) over a length $l$. The probability that the neutron remains in the spin-up state after the interaction is
\begin{equation}
    P_{\uparrow} = \cos^2\!\left(\frac{\omega_L l}{2v}\right),
\end{equation}
where $\omega_L$ is the Larmor frequency and $v$ is the neutron velocity. If the interaction region is divided into $N_Z$ segments and a spin measurement is performed after each segment, the survival probability becomes
\begin{equation}
    P_{\uparrow}^{(N_Z)} = \cos^{2N_Z}\!\left(\frac{\omega_L l}{2N_Zv}\right)
    \xrightarrow{N_Z\to\infty} 1,
\end{equation}
demonstrating the quantum Zeno effect. Neutron spin measurements can be implemented using polarized $^3$He cells, which transmit one spin state while absorbing the opposite state~\cite{gentile2017optically}. Typical $^3$He cell polarization can reach about $85~\%$.

The loop cavity naturally realizes this repeated-measurement limit. A neutron circulating in the loop passes through the same magnetic field and spin analyzer on every round trip, with the number of loops equaling $N_Z$. Only a single magnetic field region and a single $^3$He cell are required. A small DC coil placed in one arm of the loop produces a transverse magnetic field, as commonly used in neutron interferometry~\cite{werner1975observation}. For example, a $0.1$~mT field over $2.5$~cm produces a total $\pi$ spin rotation after approximately 10 loops, which can be verified by unloading the neutrons and analyzing their spin with a second polarized $^3$He cell outside the cavity.

To observe the QZE, a polarized $^3$He cell is added to the opposite arm of the loop, with its polarization aligned along the beam direction. Cells a few centimeters in size have been successfully integrated into neutron interferometers before~\cite{huber2009precision, huber2014neutron}. With the spin measurement every loop, QZE predicts that the neutron spin remains frozen in the initial $\uparrow$ state, in contrast to the full spin flip observed when only the magnetic field is present. Comparing these two configurations provides a direct test of the QZE with polarized neutrons.

Quantum Zeno dragging uses the same measurement-induced dynamics to coherently steer the quantum state. Consider again a neutron prepared in the $\uparrow$ state, circulating in the loop with only a polarized $^3$He cell present. If the polarization axis of the cell is rotated by a small angle $\delta\theta$ between successive loops, the neutron is projected onto a slightly rotated spin state. After $N_Z$ loops, the spin follows the measurement axis, achieving a total rotation of $N_Z\delta\theta$. For $\delta\theta = 1$~mrad, a $\pi$ rotation is reached after approximately 3000 loops. Unloading the neutrons and analyzing the spin demonstrates that the spin has been coherently dragged from $\uparrow$ to $\downarrow$. This mechanism is closely related to adiabatic spin rotation, but is driven purely by measurement~\cite{burgarth2022one}.

 \section{Conclusion}
We have designed and modeled a neutron loop cavity that coherently guides neutrons through a repeated closed-loop path via Bragg diffraction from perfect crystals. We estimate a confinement probability of $\sim64~\%$ for 10,000 reflections, enabling precision studies of small interactions amplified by repeated traversal of the same interaction region.

As a near-term application, we propose a next-generation neutron spin-orbit interaction measurement capable of producing a $\pi$ spin rotation within approximately 800 loop passes. This represents an order-of-magnitude improvement in sensitivity compared to recent measurements~\cite{gentile}. Moreover, an independent determination of this effect may help clarify the $\sim40~\%$ discrepancy between experimental results and theoretical predictions for the spin rotation magnitude. Relative to earlier nEDM studies using slotted crystals~\cite{dombeck2001measurement}, the loop cavity may enable a substantially increased effective interaction time. This enhancement suggests a path toward future nEDM searches with target sensitivity at the competitive $10^{-27}$~e$\cdot$cm level.

The loop cavity is also well suited for a first experimental test of the quantum Zeno effect with cold neutrons, using experimentally practical crystal sizes, magnetic field strengths, and polarized $^3$He spin filters. In addition, the cavity geometry enables parity-violating measurements with $\sim100\times$ smaller samples, while offering experimental systematics distinct from existing approaches. The loop cavity further provides a novel bottle-type geometry for neutron lifetime studies at distinct energy scales, contributing an independent data point toward resolving the neutron lifetime discrepancy.

Beyond the applications presented in this work, the neutron loop cavity offers a flexible experimental platform for exploring a range of fundamental quantum phenomena. The Aharonov-Casher effect, in which a magnetic dipole acquires a topological phase when encircling a charged electrode~\cite{aharonov1984topological}, is a natural target due to the large phase amplification achievable through repeated loop traversal. This effect was previously measured using neutron interferometry, though a scaling factor of $\sim1.5\times$ was required to reconcile experiment and theory~\cite{cimmino1989observation}. The cavity may also provide a controlled platform for studying fermionic quantum statistics. For noninteracting neutrons, the Pauli exclusion principle leads to fermion antibunching, which has been predicted and observed in thermal neutron beams~\cite{boffi1971further, silverman1988feasibility, iannuzzi2006direct, iannuzzi2011further}. Antibunching should place fundamental limits on the number of neutrons that can be simultaneously confined in the cavity.

\section*{Acknowledgments}
This work was supported by the Canadian Excellence Research Chairs (CERC) program, the Natural Sciences and Engineering Research Council of Canada (NSERC) Discovery program, the NSERC Canada Graduate Scholarships programs (CGS-M and PGS-D), the Canada  First  Research  Excellence  Fund  (CFREF), and the US Department of Energy, Office of Nuclear Physics, under Interagency Agreement 89243019SSC000025. This work was also supported by the DOE Office of Science, Office of Basic Energy Sciences, in the program ``Quantum Horizons: QIS Research and Innovation for Nuclear Science'' through grant DE-SC0023695. D.A.P and M.G.H are grateful for discussions with Muhammad Arif.

\bibliographystyle{elsarticle-num}
\bibliography{mybib.bib}

@BOOK{Sam,
  title = {{Neutron Interferometry: Lessons in Experimental Quantum Mechanics, Wave-Particle Duality, and Entanglement}},
  publisher = {Oxford University Press; 2 edition},
  year = {2015},
  volume={12},
  author = {Rauch, Helmut and Werner, Samuel A},
  isbn-13 = {978-0198712510},
}

@ARTICLE{Rauch_1975_PhysLetta,
  author = {H. Rauch and A. Zeilinger and G. Badurek and A. Wilfing and W. Bauspiess
	and U. Bonse},
  title = {Verification of coherent spinor rotation of fermions},
  journal = {Phys. Lett.},
  year = {1975},
  volume = {54A},
  pages = {425 - 427},
  number = {6},
  issn = {0375-9601},
}

@article{colella1975observation,
  title={Observation of gravitationally induced quantum interference},
  author={Colella, Roberto and Overhauser, Albert W and Werner, Samuel A},
  journal={Physical Review Letters},
  volume={34},
  number={23},
  pages={1472},
  year={1975},
  publisher={APS}
}

@article{decoherence,
  title = {Experimental Realization of Decoherence-Free Subspace in Neutron Interferometry},
  author = {Pushin, D. A. and Huber, M. G. and Arif, M. and Cory, D. G.},
  journal = {Phys. Rev. Lett.},
  volume = {107},
  issue = {15},
  pages = {150401},
  numpages = {4},
  year = {2011},
  month = {Oct},
  publisher = {American Physical Society},
}

@article{gentile,
  title = {Direct observation of neutron spin rotation in Bragg scattering due to the spin-orbit interaction in silicon},
  author = {Gentile, T. R. and Huber, M. G. and Koetke, D. D. and Peshkin, M. and Arif, M. and Dombeck, T. and Hussey, D. S. and Jacobson, D. L. and Nord, P. and Pushin, D. A. and Smither, R.},
  journal = {Phys. Rev. C},
  volume = {100},
  issue = {3},
  pages = {034005},
  numpages = {13},
  year = {2019},
  month = {Sep},
  publisher = {American Physical Society},
}

@article{cheshire_cat,
	author = {Danner, Armin and Lemmel, Hartmut and Wagner, Richard and Sponar, Stephan and Hasegawa, Yuji},
	journal = {Atoms},
	number = {6},
	title = {Neutron Interferometer Experiments Studying Fundamental Features of Quantum Mechanics},
	volume = {11},
	year = {2023}}

@article{neutron_cav, title={Quantum Information Approach to the implementation of a neutron cavity}, volume={25}, number={7}, journal={New Journal of Physics}, author={Nahman-Lévesque, O and Sarenac, D and Lailey, O and Cory, D G and Huber, M G and Pushin, D A}, year={2023}, pages={073016}}

@article{dombeck2001measurement,
  title={Measurement of the neutron reflectivity for Bragg reflections off a perfect silicon crystal},
  author={Dombeck, T. and Ringo, R. and Koetke, D. D. and Kaiser, H. and Schoen, K. and Werner, S. A. and Dombeck, D.},
  journal={Physical Review A},
  volume={64},
  number={5},
  pages={053607},
  year={2001},
  publisher={APS}
}

@article{fedorov2010measurement,
  title={Measurement of the neutron electric dipole moment via spin rotation in a non-centrosymmetric crystal},
  author={Fedorov, V. V. and Jentschel, M. and Kuznetsov, I. A. and Lapin, E. G. and Lelievre-Berna, E. and Nesvizhevsky, V. and Petoukhov, A. and Semenikhin, S. Y. and Soldner, T. and Voronin, V. V. and others},
  journal={Physics Letters B},
  volume={694},
  number={1},
  pages={22--25},
  year={2010},
  publisher={Elsevier}
}

@article{Nsofini_2016,
   title={Quantum-information approach to dynamical diffraction theory},
   volume={94},
   ISSN={2469-9934},
   number={6},
   journal={Physical Review A},
   publisher={American Physical Society (APS)},
   author={Nsofini, J. and Ghofrani, K. and Sarenac, D. and Cory, D. G. and Pushin, D. A.},
   year={2016},
   month={Dec}
}

@article{nsofini2017noise,
  title={Noise refocusing in a five-blade neutron interferometer},
  author={Nsofini, J. and Sarenac, D. and Ghofrani, K. and Huber, M. G. and Arif, M. and Cory, D. G. and Pushin, D. A.},
  journal={Journal of Applied Physics},
  volume={122},
  number={5},
  pages={054501},
  year={2017},
  publisher={AIP Publishing LLC}
}

@article{nsofini2019coherence,
  title={Coherence optimization in neutron interferometry through defocusing},
  author={Nsofini, J. and Sarenac, D. and Cory, D. G. and Pushin, D. A.},
  journal={Physical Review A},
  volume={99},
  number={4},
  pages={043614},
  year={2019},
  publisher={APS}
}

@article{nahman2022generalizing,
  title={Generalizing the quantum information model for dynamic diffraction},
  author={Nahman-L{\'e}vesque, O. and Sarenac, D. and Cory, D. G. and Heacock, B. and Huber, M. G. and Pushin, D. A.},
  journal={Physical Review A},
  volume={105},
  number={2},
  pages={022403},
  year={2022},
  publisher={APS}
}

@article{vesta,
	author = {M.R. Jaekel and E. Jericha and H. Rauch},
	journal = {Nuclear Instruments and Methods in Physics Research Section A: Accelerators, Spectrometers, Detectors and Associated Equipment},
	number = {1},
	pages = {335-344},
	title = {New developments in cold neutron storage with perfect crystals},
	volume = {539},
	year = {2005}}

@article{BB_storage,
	author = {E. Jericha and C.J. Carlile and M. J{\"a}kel and H. Rauch},
	journal = {Physica B: Condensed Matter},
	pages = {1066-1067},
	title = {Cold neutron storage by perfect crystals},
	volume = {234-236},
	year = {1997}}

@article{huber2024achieving,
  title={Achieving a near-ideal silicon crystal neutron interferometer using submicrometer fabrication techniques},
  author={Huber, MG and Heacock, B and Taminiau, I and Cory, DG and Sarenac, D and Valdillez, R and Pushin, DA},
  journal={Physical Review Research},
  volume={6},
  number={4},
  pages={043079},
  year={2024},
  publisher={APS}
}

@article{schuster1992cold,
  title={A cold neutron storage device and a neutron resonator},
  author={Schuster, M and Jericha, E and Carlile, CJ and Rauch, H},
  journal={Physica B: Condensed Matter},
  volume={180},
  pages={997--999},
  year={1992},
  publisher={Elsevier}
}

@article{jericha1996performance,
  title={Performance of an improved perfect crystal neutron storage cavity},
  author={Jericha, E and Carlile, CJ and Rauch, H},
  journal={Nuclear Instruments and Methods in Physics Research Section A: Accelerators, Spectrometers, Detectors and Associated Equipment},
  volume={379},
  number={2},
  pages={330--334},
  year={1996},
  publisher={Elsevier}
}

@book{authier2012x,
  title={X-ray and neutron dynamical diffraction: theory and applications},
  author={Authier, Andr{\'e} and Lagomarsino, Stefano and Tanner, Brian K},
  volume={357},
  year={2012},
  publisher={Springer Science \& Business Media}
}

@article{facchi2003optimization,
  title={Optimization of a neutron-spin test of the quantum Zeno effect},
  author={Facchi, Paolo and Nakaguro, Yoichi and Nakazato, Hiromichi and Pascazio, Saverio and Unoki, Makoto and Yuasa, Kazuya},
  journal={Physical Review A},
  volume={68},
  number={1},
  pages={012107},
  year={2003},
  publisher={APS}
}

@inproceedings{jaekel1999new,
  title={New measurements with a perfect crystal cavity for neutrons},
  author={Jaekel, Martin R and Carlile, Colin J and Jericha, Erwin and Schwab, Dagmar E and Rauch, Helmut},
  booktitle={EUV, X-Ray, and Neutron Optics and Sources},
  volume={3767},
  pages={353--359},
  year={1999},
  organization={SPIE}
}

@article{schwinger1948polarization,
  title={On the polarization of fast neutrons},
  author={Schwinger, Julian},
  journal={Physical Review},
  volume={73},
  number={4},
  pages={407},
  year={1948},
  publisher={APS}
}

@article{shull1967search,
  title={Search for a neutron electric dipole moment by a scattering experiment},
  author={Shull, CG and Nathans, R},
  journal={Physical Review Letters},
  volume={19},
  number={7},
  pages={384},
  year={1967},
  publisher={APS}
}

@article{shull1963neutron,
  title={Neutron spin-neutron orbit interaction with slow neutrons},
  author={Shull, CG},
  journal={Physical Review Letters},
  volume={10},
  number={7},
  pages={297},
  year={1963},
  publisher={APS}
}

@inproceedings{fedorov2018modern,
  title={Modern Status of Searches for nEDM, Using Neutron Optics and Diffraction in Noncentrosymmetric Crystals},
  author={Fedorov, VV and Voronin, VV},
  booktitle={Proceedings of the International Conference on Neutron Optics (NOP2017)},
  pages={011007},
  year={2018}
}

@article{fedorov2006neutron,
  title={Neutron spin optics in noncentrosymmetric crystals as a new way for nEDM search},
  author={Fedorov, VV and Kuznetsov, IA and Lapin, EG and Semenikhin, S Yu and Voronin, VV},
  journal={Nuclear Instruments and Methods in Physics Research Section B: Beam Interactions with Materials and Atoms},
  volume={252},
  number={1},
  pages={131--135},
  year={2006},
  publisher={Elsevier}
}

@article{lamoreaux2009experimental,
  title={Experimental searches for the neutron electric dipole moment},
  author={Lamoreaux, SK and Golub, R},
  journal={Journal of Physics G: Nuclear and Particle Physics},
  volume={36},
  number={10},
  pages={104002},
  year={2009},
  publisher={IOP Publishing}
}

@article{abel2020measurement,
  title={Measurement of the permanent electric dipole moment of the neutron},
  author={Abel, Christopher and Afach, Samer and Ayres, Nicholas J and Baker, Colin A and Ban, Gilles and Bison, Georg and Bodek, Kazimierz and Bondar, Vira and Burghoff, Martin and Chanel, Estelle and others},
  journal={Physical Review Letters},
  volume={124},
  number={8},
  pages={081803},
  year={2020},
  publisher={APS}
}

@article{pascazio1993quantum,
  title={Quantum Zeno effect with neutron spin},
  author={Pascazio, Saverio and Namiki, Mikio and Badurek, Gerald and Rauch, Helmut},
  journal={Physics Letters A},
  volume={179},
  number={3},
  pages={155--160},
  year={1993},
  publisher={Elsevier}
}

@article{misra1977zeno,
  title={The Zeno’s paradox in quantum theory},
  author={Misra, Baidyanath and Sudarshan, EC George},
  journal={Journal of Mathematical Physics},
  volume={18},
  number={4},
  pages={756--763},
  year={1977},
  publisher={American Institute of Physics}
}

@article{itano1990quantum,
  title={Quantum zeno effect},
  author={Itano, Wayne M and Heinzen, Daniel J and Bollinger, John J and Wineland, David J},
  journal={Physical Review A},
  volume={41},
  number={5},
  pages={2295},
  year={1990},
  publisher={APS}
}

@article{rauch2001quantum,
  title={Quantum Zeno-effect with polarized neutrons},
  author={Rauch, H},
  journal={Physica B: Condensed Matter},
  volume={297},
  number={1-4},
  pages={299--302},
  year={2001},
  publisher={Elsevier}
}

@article{li2016neutron,
  title={Neutron limit on the strongly-coupled chameleon field},
  author={Li, Ke and Arif, Muhammad and Cory, David G and Haun, Robert and Heacock, Benjamin and Huber, Michael G and Nsofini, Joachim and Pushin, Dimitry A and Saggu, Parminder and Sarenac, Dusan and others},
  journal={Physical Review D},
  volume={93},
  number={6},
  pages={062001},
  year={2016},
  publisher={APS}
}

@article{zeilinger1984symmetry,
  title={Symmetry violations and Schwinger scattering in neutron interferometry},
  author={Zeilinger, A and Horne, MA and Bernstein, HJ},
  journal={Le Journal de Physique Colloques},
  volume={45},
  number={C3},
  pages={C3--209},
  year={1984},
  publisher={EDP Sciences}
}

@article{heacock2021pendellosung,
  title={Pendell{\"o}sung interferometry probes the neutron charge radius, lattice dynamics, and fifth forces},
  author={Heacock, Benjamin and Fujiie, Takuhiro and Haun, Robert W and Henins, Albert and Hirota, Katsuya and Hosobata, Takuya and Huber, Michael G and Kitaguchi, Masaaki and Pushin, Dmitry A and Shimizu, Hirohiko and others},
  journal={Science},
  volume={373},
  number={6560},
  pages={1239--1243},
  year={2021},
  publisher={American Association for the Advancement of Science}
}

@article{clark2015controlling,
  title={Controlling neutron orbital angular momentum},
  author={Clark, Charles W and Barankov, Roman and Huber, Michael G and Arif, Muhammad and Cory, David G and Pushin, Dmitry A},
  journal={Nature},
  volume={525},
  number={7570},
  pages={504--506},
  year={2015},
  publisher={Nature Publishing Group UK London}
}

@article{sarenac2016holography,
  title={Holography with a neutron interferometer},
  author={Sarenac, Dusan and Huber, Michael G and Heacock, Benjamin and Arif, Muhammad and Clark, Charles W and Cory, David G and Shahi, Chandra B and Pushin, Dmitry A},
  journal={Optics express},
  volume={24},
  number={20},
  pages={22528--22535},
  year={2016},
  publisher={Optical Society of America}
}

@article{geerits2023phase,
  title={Phase vortex lattices in neutron interferometry},
  author={Geerits, Niels and Lemmel, Hartmut and Berger, Anna-Sophie and Sponar, Stephan},
  journal={Communications Physics},
  volume={6},
  number={1},
  pages={209},
  year={2023},
  publisher={Nature Publishing Group UK London}
}

@article{schuster1990test,
  title={Test of a perfect crystal neutron storage device},
  author={Schuster, M and Rauch, H and Seidl, E and Jericha, E and Carlile, CJ},
  journal={Physics Letters A},
  volume={144},
  number={6-7},
  pages={297--300},
  year={1990},
  publisher={Elsevier}
}

@article{werner1975observation,
  title={Observation of the phase shift of a neutron due to precession in a magnetic field},
  author={Werner, Samuel A and Colella, Roberto and Overhauser, Albert W and Eagen, CF},
  journal={Physical Review Letters},
  volume={35},
  number={16},
  pages={1053},
  year={1975},
  publisher={APS}
}

@article{huber2009precision,
  title={Precision Measurement of the n-He 3 Incoherent Scattering Length Using Neutron Interferometry},
  author={Huber, Michael G and Arif, Muhammad and Black, TC and Chen, WC and Gentile, Thomas R and Hussey, Daniel S and Pushin, DA and Wietfeldt, <? format?> FE and Yang, Liang},
  journal={Physical review letters},
  volume={102},
  number={20},
  pages={200401},
  year={2009},
  publisher={APS}
}

@article{huber2014neutron,
  title={Neutron interferometric measurement of the scattering length difference between the triplet and singlet states of n-He 3},
  author={Huber, Michael G and Arif, Muhammad and Chen, Wangchun C and Gentile, Thomas R and Hussey, Daniel S and Black, Timothy C and Pushin, Dimitry A and Shahi, Chandra B and Wietfeldt, Fred E and Yang, Liang},
  journal={Physical Review C},
  volume={90},
  number={6},
  pages={064004},
  year={2014},
  publisher={APS}
}

@article{iannuzzi2006direct,
  title={Direct experimental evidence of free-fermion antibunching},
  author={Iannuzzi, M and Orecchini, Andrea and Sacchetti, Francesco and Facchi, Paolo and Pascazio, Saverio},
  journal={Physical review letters},
  volume={96},
  number={8},
  pages={080402},
  year={2006},
  publisher={APS}
}

@article{iannuzzi2011further,
  title={Further evidence of antibunching of two coherent beams of fermions},
  author={Iannuzzi, M and Messi, R and Moricciani, D and Orecchini, Andrea and Sacchetti, Francesco and Facchi, Paolo and Pascazio, Saverio},
  journal={Physical Review A—Atomic, Molecular, and Optical Physics},
  volume={84},
  number={1},
  pages={015601},
  year={2011},
  publisher={APS}
}

@article{silverman1988feasibility,
  title={On the feasibility of a neutron Hanbury Brown—Twiss experiment with gravitationally-induced phase shift},
  author={Silverman, MP},
  journal={Physics Letters A},
  volume={132},
  number={4},
  pages={154--158},
  year={1988},
  publisher={Elsevier}
}

@article{boffi1971further,
  title={Further remarks on the coherence properties of a thermal neutron beam},
  author={Boffi, S and Caglioti, G},
  journal={Il Nuovo Cimento B (1971-1996)},
  volume={3},
  number={2},
  pages={262--268},
  year={1971},
  publisher={Springer}
}

@article{musedinovic2025measurement,
  title={Measurement of the free neutron lifetime in a magneto-gravitational trap with in situ detection},
  author={Musedinovic, R and Blokland, L\_S and Cude-Woods, C\_B and Singh, M and Blatnik, M\_A and Callahan, N and Choi, J\_H and Clayton, S\_M and Filippone, B\_W and Fox, W\_R and others},
  journal={Physical Review C},
  volume={111},
  number={4},
  pages={045501},
  year={2025},
  publisher={APS}
}

@article{caylor2025detection,
  title={Detection of Molecular Hydrogen in a Proton Trap},
  author={Caylor, J and Biswas, R and Crawford, B and Dewey, MS and Fomin, N and Greene, GL and Hoogerheide, SF and Hungria-Negron, J and Mumm, HP and Nico, JS and others},
  journal={arXiv preprint arXiv:2506.01682},
  year={2025}
}

@article{fuwa2024improved,
  title={Improved measurements of neutron lifetime with cold neutron beam at J-PARC},
  author={Fuwa, Y and Hasegawa, T and Hirota, K and Hoshino, T and Hosokawa, R and Ichikawa, G and Ieki, S and Ino, T and Iwashita, Y and Kitaguchi, M and others},
  journal={arXiv preprint arXiv:2412.19519},
  year={2024}
}

@article{chowdhury2023neutron,
  title={Neutron lifetime anomaly and big bang nucleosynthesis},
  author={Chowdhury, Tammi and Ipek, Seyda},
  journal={Canadian Journal of Physics},
  volume={102},
  number={2},
  pages={96--99},
  year={2023},
  publisher={Canadian Science Publishing 1840 Woodward Drive, Suite 1, Ottawa, ON K2C 0P7}
}

@article{snow2011upper,
  title={Upper bound on parity-violating neutron spin rotation in He 4},
  author={Snow, William Michael and Bass, CD and Bass, TD and Crawford, BE and Gan, K and Heckel, BR and Luo, D and Markoff, DM and Micherdzinska, AM and Mumm, HP and others},
  journal={Physical Review C—Nuclear Physics},
  volume={83},
  number={2},
  pages={022501},
  year={2011},
  publisher={APS}
}

@article{fischer2001observation,
  title={Observation of the quantum Zeno and anti-Zeno effects in an unstable system},
  author={Fischer, Martin C and Guti{\'e}rrez-Medina, Braulio and Raizen, Mark G},
  journal={Physical review letters},
  volume={87},
  number={4},
  pages={040402},
  year={2001},
  publisher={APS}
}

@article{kwiat1995interaction,
  title={Interaction-free measurement},
  author={Kwiat, Paul and Weinfurter, Harald and Herzog, Thomas and Zeilinger, Anton and Kasevich, Mark A},
  journal={Physical Review Letters},
  volume={74},
  number={24},
  pages={4763},
  year={1995},
  publisher={APS}
}

@article{lewalle2024optimal,
  title={Optimal zeno dragging for quantum control: a shortcut to zeno with action-based scheduling optimization},
  author={Lewalle, Philippe and Zhang, Yipei and Whaley, K Birgitta},
  journal={PRX Quantum},
  volume={5},
  number={2},
  pages={020366},
  year={2024},
  publisher={APS}
}

@article{hacohen2018incoherent,
  title={Incoherent qubit control using the quantum Zeno effect},
  author={Hacohen-Gourgy, Shay and Garc{\'\i}a-Pintos, Luis Pedro and Martin, Leigh S and Dressel, Justin and Siddiqi, Irfan},
  journal={Physical review letters},
  volume={120},
  number={2},
  pages={020505},
  year={2018},
  publisher={APS}
}

@article{burgarth2022one,
  title={One bound to rule them all: from Adiabatic to Zeno},
  author={Burgarth, Daniel and Facchi, Paolo and Gramegna, Giovanni and Yuasa, Kazuya},
  journal={Quantum},
  volume={6},
  pages={737},
  year={2022},
  publisher={Verein zur F{\"o}rderung des Open Access Publizierens in den Quantenwissenschaften}
}

@article{cimmino1989observation,
  title={Observation of the topological Aharonov-Casher phase shift by neutron interferometry},
  author={Cimmino, A and Opat, GI and Klein, AG and Kaiser, H and Werner, SA and Arif, M and Clothier, R},
  journal={Physical review letters},
  volume={63},
  number={4},
  pages={380},
  year={1989},
  publisher={APS}
}

@article{aharonov1984topological,
  title={Topological quantum effects for neutral particles},
  author={Aharonov, Yakir and Casher, Aharon},
  journal={Physical Review Letters},
  volume={53},
  number={4},
  pages={319},
  year={1984},
  publisher={APS}
}

@article{bonse1968two,
  title={A two-crystal X-ray interferometer},
  author={Bonse, U and Te Kaat, E},
  journal={Zeitschrift f{\"u}r Physik A Hadrons and nuclei},
  volume={214},
  number={1},
  pages={16--21},
  year={1968},
  publisher={Springer}
}

@article{lemmel2022neutron,
  title={Neutron interference from a split-crystal interferometer},
  author={Lemmel, Hartmut and Jentschel, Michael and Abele, Hartmut and Lafont, Fabien and Guerard, Bruno and Sasso, Carlo P and Mana, Giovanni and Massa, Enrico},
  journal={Applied Crystallography},
  volume={55},
  number={4},
  pages={870--875},
  year={2022},
  publisher={International Union of Crystallography}
}

@article{deslattes1968x,
  title={X-ray monochromators and resonators from single crystals},
  author={Deslattes, Richard D},
  journal={Applied Physics Letters},
  volume={12},
  number={4},
  pages={133--135},
  year={1968}
}

@article{bond1967proposed,
  title={Proposed resonator for an X-ray laser},
  author={Bond, WL and Duguay, MA and Rentzepis, PM},
  journal={Applied Physics Letters},
  volume={10},
  number={8},
  pages={216--218},
  year={1967}
}

@article{cotterill1968universal,
  title={A universal planar X-ray resonator},
  author={Cotterill, RMJ},
  journal={Applied Physics Letters},
  volume={12},
  number={12},
  pages={403--404},
  year={1968}
}

@article{liu2024x,
  title={X-ray optics for the cavity-based X-ray free-electron laser},
  author={Liu, Peifan and Pradhan, Paresh and Shi, Xianbo and Shu, Deming and Kauchha, Keshab and Qiao, Zhi and Tamasaku, Kenji and Osaka, Taito and Zhu, Diling and Sato, Takahiro and others},
  journal={Synchrotron Radiation},
  volume={31},
  number={4},
  pages={751--762},
  year={2024},
  publisher={International Union of Crystallography}
}

@article{schindler2013theory,
  title={The theory of parity violation in few-nucleon systems},
  author={Schindler, Matthias R and Springer, Roxanne P},
  journal={Progress in Particle and Nuclear Physics},
  volume={72},
  pages={1--43},
  year={2013},
  publisher={Elsevier}
}

@article{gentile2017optically,
  title={Optically polarized he 3},
  author={Gentile, Thomas R and Nacher, PJ and Saam, B and Walker, TG},
  journal={Reviews of modern physics},
  volume={89},
  number={4},
  pages={045004},
  year={2017},
  publisher={APS}
}

@article{bass2009liquid,
  title={A liquid helium target system for a measurement of parity violation in neutron spin rotation},
  author={Bass, Christopher D and Bass, Tiffany D and Heckel, BR and Huffer, CR and Luo, D and Markoff, DM and Micherdzinska, AM and Snow, WM and Swanson, HE and Walbridge, SC},
  journal={Nuclear Instruments and Methods in Physics Research Section A: Accelerators, Spectrometers, Detectors and Associated Equipment},
  volume={612},
  number={1},
  pages={69--82},
  year={2009},
  publisher={Elsevier}
}

@article{le2023upgrade,
  title={Upgrade of the MARI spectrometer at ISIS},
  author={Le, Manh Duc and Guidi, Tatiana and Bewley, R and Stewart, J Ross and Schooneveld, Erik M and Raspino, Davide and Pooley, Daniel E and Boxall, Jonathan and Gascoyne, Kelvin F and Rhodes, Nigel J and others},
  journal={Nuclear Instruments and Methods in Physics Research Section A: Accelerators, Spectrometers, Detectors and Associated Equipment},
  volume={1056},
  pages={168646},
  year={2023},
  publisher={Elsevier}
}

@misc{nist_guide_upgrade,
  author       = {{National Institute of Standards and Technology}},
  title        = {{NCNR} Guide Upgrade},
  year         = {2026},
  note         = {{NIST} Center for Neutron Research},
  urldate      = {2026-03-11}
}

@article{abele2006characterization,
  title={Characterization of a ballistic supermirror neutron guide},
  author={Abele, Hartmut and Dubbers, D and H{\"a}se, H and Klein, M and Kn{\"o}pfler, A and Kreuz, M and Lauer, T and M{\"a}rkisch, B and Mund, D and Nesvizhevsky, V and others},
  journal={Nuclear Instruments and Methods in Physics Research Section A: Accelerators, Spectrometers, Detectors and Associated Equipment},
  volume={562},
  number={1},
  pages={407--417},
  year={2006},
  publisher={Elsevier}
}
    
\end{document}